\documentclass[%
 reprint,
 amsmath,amssymb,
 aps,
]{revtex4-1}

\usepackage{graphicx}% Include figure files
\usepackage{dcolumn}% Align table columns on decimal point
\usepackage{bm}% bold math
\usepackage{nicefrac}
\usepackage{upgreek}

\begin{document}

\preprint{APS/123-QED}

\title{Cryogenic second harmonic generation\\in periodically-poled lithium niobate waveguides}
\author{Moritz Bartnick$^{1,*}$, Matteo Santandrea$^{2}$, Jan Philipp Höpker$^{1}$, Frederik Thiele$^{1}$, Raimund Ricken$^{2}$, Viktor Quiring$^{2}$, Christof Eigner$^{2}$, Harald Herrmann$^{2}$, Christine Silberhorn$^{2}$ and Tim J. Bartley$^{1}$}
\affiliation{
\vspace{1mm}$^1$~Mesoscopic Quantum Optics, Department of Physics, Paderborn University,\\Warburger Str. 100, 33098 Paderborn, Germany
}
\affiliation{\vspace{1mm}
$^2$~Integrated Quantum Optics, Department of Physics, Paderborn University,\\Warburger Str. 100, 33098 Paderborn, Germany
}
\affiliation{\vspace{1mm}
$^{*}$ bartnick@mail.uni-paderborn.de\vspace{2mm}
}

\date{\today}% It is always \today, today,
             %  but any date may be explicitly specified

\begin{abstract}
Prospective integrated quantum optical technologies will combine nonlinear optics and components requiring cryogenic operating temperatures. Despite the prevalence of integrated platforms exploiting $\chi^{(2)}$-nonlinearities for quantum optics, for example used for quantum state generation and frequency conversion, their material properties at low temperatures are largely unstudied. Here, we demonstrate the first second harmonic generation in a fiber-coupled lithium niobate waveguide at temperatures down to 4.4\,K. We observe a reproducible shift in the phase-matched pump wavelength within the telecom band, in addition to transient discontinuities while temperature cycling. Our results establish lithium niobate as a versatile nonlinear photonic integration platform compatible with cryogenic quantum technologies.
\end{abstract}

%\keywords{Suggested keywords}%Use showkeys class option if keyword
                              %display desired
\maketitle

%\tableofcontents

\section{Introduction}
Integrated photonics has become the standard implementation for many quantum optical technologies~\cite{wang2019}. However, it is becoming increasingly apparent that no single material platform will achieve optimal generation, manipulation and detection of quantum states simultaneously. Thus, hybrid solutions are gaining significant interest~\cite{kurizki2015,bogdanov2017,kim2020}, since they combine the optimal approaches to different tasks. For example, single-photon sources based on semiconductor quantum dots, manipulation by electro-optic switching and frequency conversion in nonlinear media, as well as detection by thin-film superconductors, are currently the leading approaches individually. Although the required materials can generally be mutually integrated, achieving functionality under common operating conditions must be ensured. For example, high-performance single-photon emitters and superconducting detectors require cryogenic operating temperatures in vacuum, whereas frequency conversion and switching is typically achieved under ambient conditions. Raising the operating temperatures of sources and detectors is likely impossible whilst maintaining their optimal performance. It is therefore crucial to explore switching and frequency conversion platforms at low temperatures to enable mutual compatibility.

%%%%
Among the diverse technological platforms for implementing switching and frequency conversion, lithium niobate is an excellent candidate, since it exhibits a large electro-optic coefficient and high second order optical nonlinearity. Moreover, it is a very mature technology which underpins much of the modern telecommunication network infrastructure. The properties of bulk as well as integrated devices in lithium niobate under ambient conditions are well-understood and their performance has been optimized accordingly, thanks to extensive academic and commercial development for almost 50 years~\cite{lawrence1993,sohler2008,sharapova2017}. 
%%%%

%When considering lithium niobate as a platform to build a hybrid, quantum-compatible device, integration within a larger quantum network necessitates the use of waveguides. 
There exists many implementations of waveguides in lithium niobate~\cite{digonnet1985,volk2016,wang2017}, of which titanium in-diffusion is an excellent candidate for cryogenic-compatible quantum protocols. It offers both low-loss propagation~\cite{sharapova2017} and highly efficient coupling to single-mode fibers in the telecom band~\cite{montaut2017}, as well as being compatible with superconducting detectors~\cite{tanner2012,hoepker2019}. Furthermore, when combined with periodic poling, such waveguides provide the modal confinement necessary for high-efficiency nonlinear interactions. Despite these advantages, there exists relatively few studies of nonlinear optical phenomena in lithium niobate in a cryogenic environment~\cite{huang2013,carbajo2015}, since demand for operation under these conditions has been limited until now.
%However, such interactions have not been extensively studied at low temperatures.

%%%%
In this paper, we demonstrate the first quasi-phase-matched second harmonic generation (SHG) in periodically-poled, titanium in-diffused waveguides in congruently-grown lithium niobate at temperatures down to 4.4\,K. We investigate the temperature-dependence of the SHG spectrum, which enables us to study how temperature changes affect the material properties from 292\,K down to 4.4\,K. We observe deviations from current models of the Sellmeier equations, as well as temperature-dependent dynamics attributable to the pyroelectric effect in lithium niobate. %To the best of our knowledge, our results represent the first observations of cryogenic SHG in integrated nonlinear circuits, which are fully compatible with, and have been extensively used for, quantum photonic technologies~\cite{sharapova2017,luo2019}.

\section{Temperature dependence of Second Harmonic Generation}
%We investigated type-II SHG in periodically-poled titanium in-diffused lithium niobate waveguides (Ti:PPLN waveguides) from the C-band at cryogenic temperatures. 

The nonlinear interaction we implement is Type-II SHG, whereby two electromagnetic waves in orthogonally-polarized waveguide modes and with equal wavelengths $\lambda_\mathrm{p}$, combine to produce a third wave with half the wavelength $\lambda_\mathrm{p}/2$. The power of the generated second harmonic light $P_\textrm{SHG}$ depends on the effective second-order susceptibility tensor element $d_\textrm{eff}$, %the pump intensity $I_\textrm{p}$, 
as well as the phase-mismatch of the interacting fields $\Delta \beta$ and the sample length $L$, and scales as~\cite{boyd2008}
\begin{equation}\label{eq:sinc}
P_\mathrm{SHG}\propto d_\mathrm{eff}^2\mathrm{sinc}^2 \left(\frac{\Delta \beta L}{2}\right)~,
\end{equation}
where $d_{13}$ is the relevant second-order susceptibility tensor element for Type-II phase-matching. While other configurations  would enable a stronger interaction (for example Type-0 phase-matching exploits the large $d_{33}$ element) we choose Type-II phase-matching to yield insight into other nonlinear processes which use $d_{13}$, such as degenerate parametric down-conversion in orthogonal polarisation modes~\cite{montaut2017}.

From Eq.~\ref{eq:sinc}, maximal conversion is achieved when $\Delta \beta=0$, which occurs at the phase-matched wavelength $\bar{\lambda}_p$.
For our process, the phase-mismatch can be expressed by
\begin{align}\label{eq:delta_k_q}
%\Delta\beta=2 \pi\bigg[&\frac{2n_\mathrm{TE}(\nicefrac{\lambda_\mathrm{p}}{2},T)-n_\mathrm{TM}(\lambda_\mathrm{p},T)-n_\mathrm{TE}(\lambda_\mathrm{p},T)}{\lambda_\mathrm{p}}\nonumber\\
%&-\frac{1}{\Lambda(T)}\bigg],
\frac{\Delta\beta}{2 \pi}=\frac{2n_\mathrm{TE}(\nicefrac{\lambda_\mathrm{p}}{2},T)-n_\mathrm{TM}(\lambda_\mathrm{p},T)-n_\mathrm{TE}(\lambda_\mathrm{p},T)}{\lambda_\mathrm{p}}-\frac{1}{\Lambda(T)},
\end{align}
where $T$ designates the waveguide temperature and $\Lambda(T)$ the poling period. Temperature dependence of $\Delta\beta$ arises due to thermal expansion of the poling period, as well as the temperature-dependent effective refractive indices of the interacting waves $n_\mathrm{TE}(\nicefrac{\lambda_\mathrm{p}}{2},T)$, $n_\mathrm{TM}(\lambda_\mathrm{p},T)$ and $n_\mathrm{TE}(\lambda_\mathrm{p},T)$. Above 300\,K, empiric Sellmeier relations for the refractive indices of congruent lithium niobate are known~\cite{edwards1984,jundt1997}. In our experiment, we explore the behavior of the phase-matching relation in \eqref{eq:delta_k_q} below 300\,K.

\section{Experimental setup}
\begin{figure}[tp]%<left> <lower> <right> <upper>
\centering\includegraphics[width=\linewidth,trim={2.5cm 11cm 2.0cm 24cm},clip]{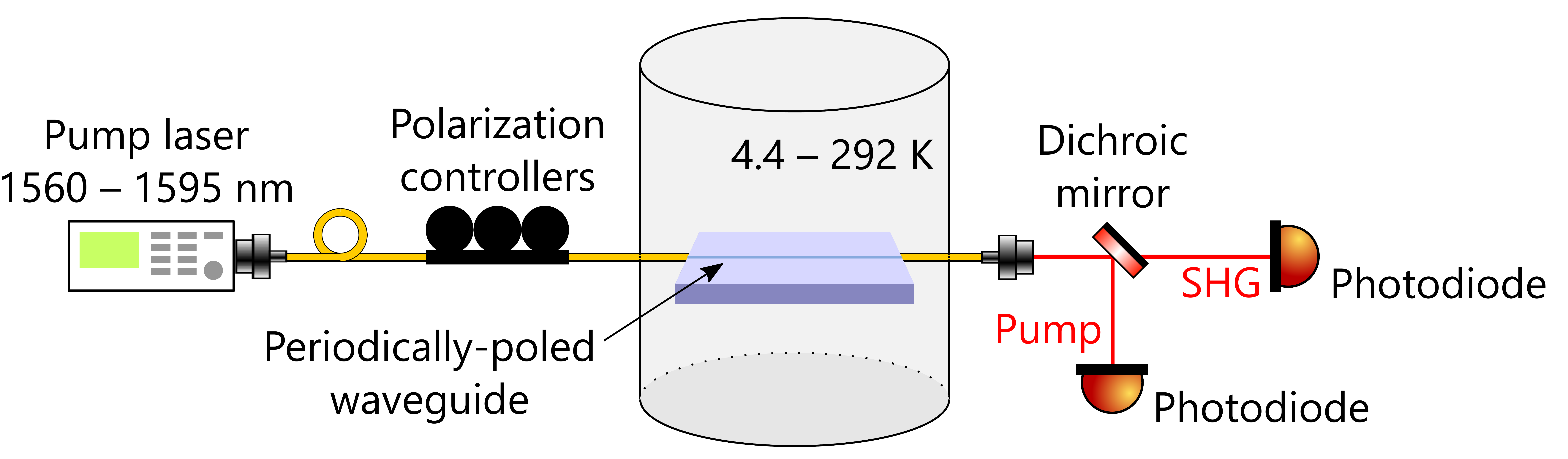}
\caption{Experimental setup to measure SHG spectra from a fiber-pigtailed waveguide sample inside a cryostat. All other components are at room temperature. For details see text.}\label{fig:experimental_setup}
\end{figure}

The sample is fabricated by in-diffusing a titanium strip of width $7\,\mathrm{\upmu m}$ and height 80\,nm into a 25\,mm-long $z$-cut congruently-grown lithium niobate chip. Using the electric-field poling method~\cite{yamada1993}, the waveguide is poled with a period of $\Lambda=9.12\,\mathrm{\upmu m}$, such that at room temperature, Type-II SHG is phase-matched for a pump wavelength in the telecom C-band. To enable robust low-temperature operation, the chip is fiber-coupled (``pigtailed'') by gluing optical single-mode fibers to the waveguide end-facets~\cite{montaut2017,hoepker2019}.

For the low-temperature measurements, the lithium niobate chip is placed inside a Gifford McMahon cryocooler and the pigtailed fibers are spliced to optical fiber feedthroughs (see Fig.~\ref{fig:experimental_setup}). We pump the nonlinear waveguide with a tunable narrow-band CW infrared laser ($\ll 1\,\mathrm{pm}$ FWHM). At the feedthrough into the cryostat, a total pump power of 17.5\,mW is incident and we set the waveguide input polarization such that the SHG signal is maximized. The pump and second harmonic light are coupled out and split by a dichroic mirror. We measure the SHG power and pump transmission with amplified photodiodes. During temperature cycling between 4.4\,K and 292\,K, the  transmission varied between 17~--~44\,\% due to thermal contractions causing misalignment of the waveguide-to-fiber interconnections. Nevertheless, SHG is clearly observed throughout the entire temperature range, with a conversion efficiency varying between $(0.04\substack{+0.05 \\ -0.02})\,\%\mathrm{W}^{-1}\mathrm{cm}^{-2}$ at 4.4\,K and $(0.10\substack{+0.10 \\ -0.05})\,\%\mathrm{W}^{-1}\mathrm{cm}^{-2}$ at 292\,K. The error bars arise from the uncertainty in the individual transmissions through both waveguide facets; further details are provided in App.~\ref{sec:conv_eff_app}.

%We observe the phase-matching of the SHG process while the sample is cooled down to $4.4\,\mathrm{K}$ ($\sim$15\,h duration) and subsequently warmed up ($\sim$20\,h duration). Over a time period of 260\,s, we continuously measure SHG spectra by sweeping the pump wavelength between 1560\,nm and 1595\,nm in steps of 0.1\,nm. 

%For the experimental data presented in this manuscript, the sample was cooled down and subsequently warmed up twice. 

\section{Cryogenic SHG Spectra}
\begin{figure}[tb]
	\centering %l, b, r, t
		\includegraphics[width=\linewidth, trim =0mm 0mm 0mm 3mm, clip]{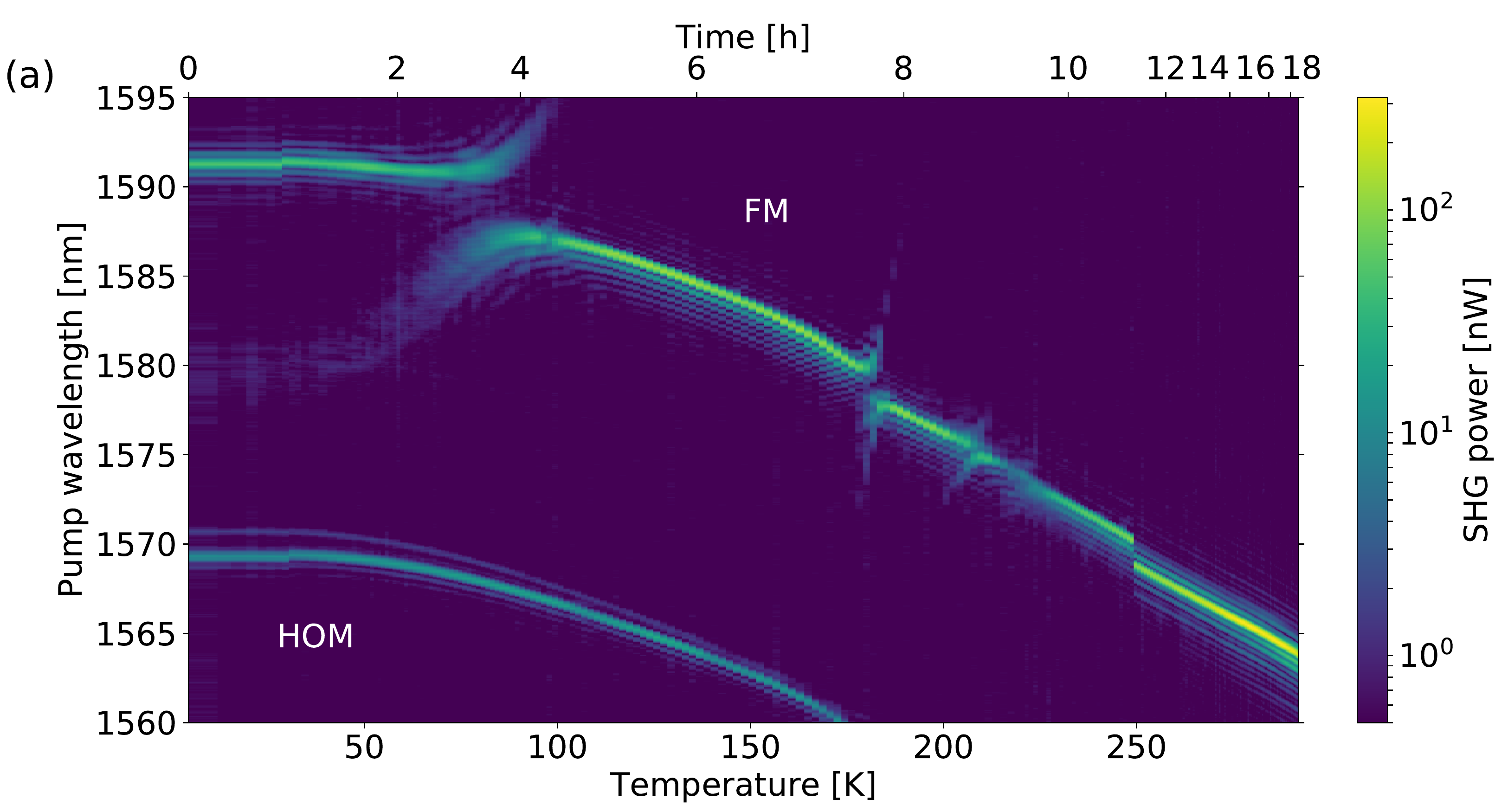}
		\includegraphics[width=\linewidth, trim =0mm 0mm 0mm 8.5mm, clip]{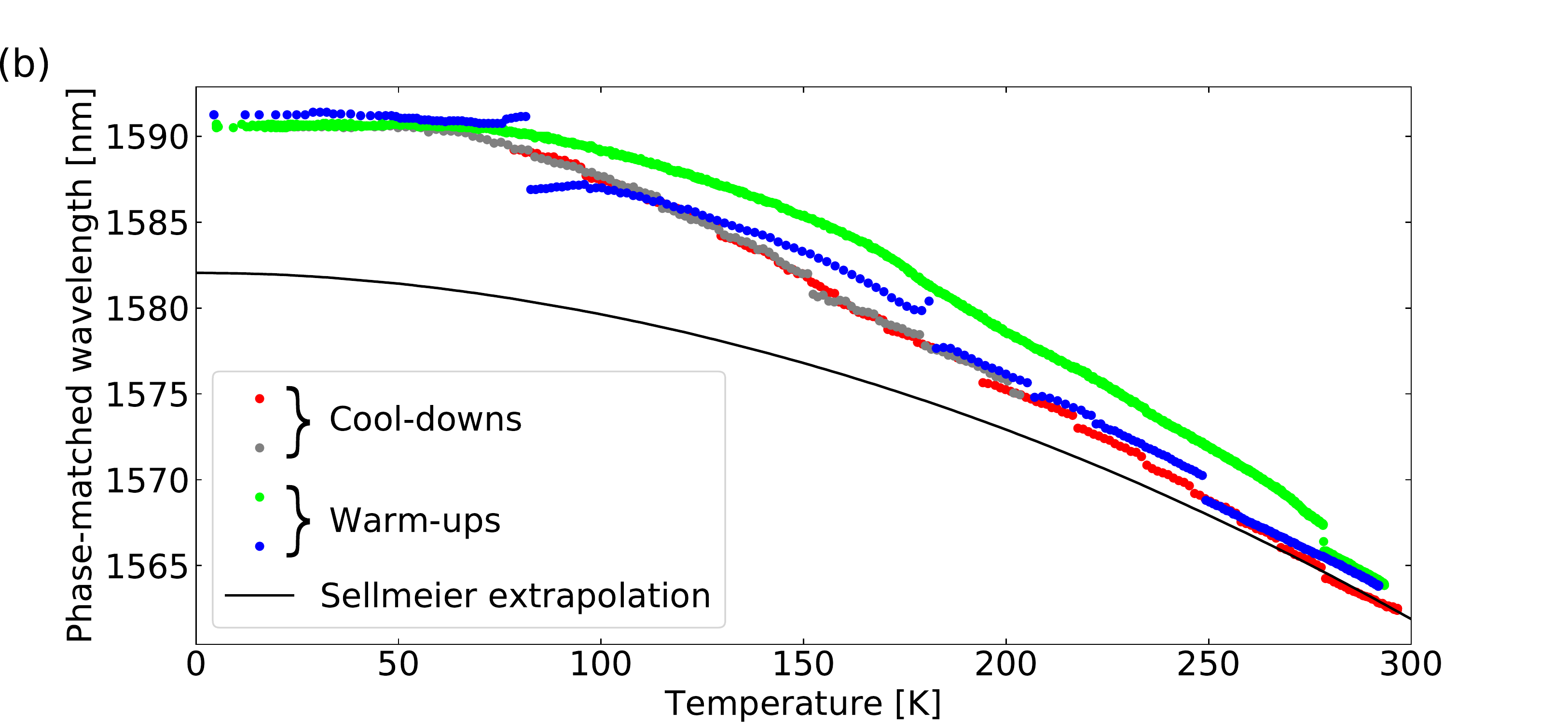}
	\caption{(a) Measured SHG power dependent on pump wavelength and temperature, in the fundamental (FM) and higher-order mode (HOM).    (b) Phase-matched pump wavelength $\bar{\lambda}_p$ of the fundamental mode as a function of temperature. The curve calculated from the extrapolated Sellmeier equations (black line) predicts a smaller wavelength shift than that which is measured in two separate temperature cycles (red, gray dots: cool-downs; blue, green dots: warm-ups).}
	\label{fig:experimental_data}
\end{figure}
\begin{figure*}[tb]
	\centering %<left> <lower> <right> <upper>
		\includegraphics[width=\linewidth,trim={2.4cm 0.2cm 2.6cm 0.7cm},clip]{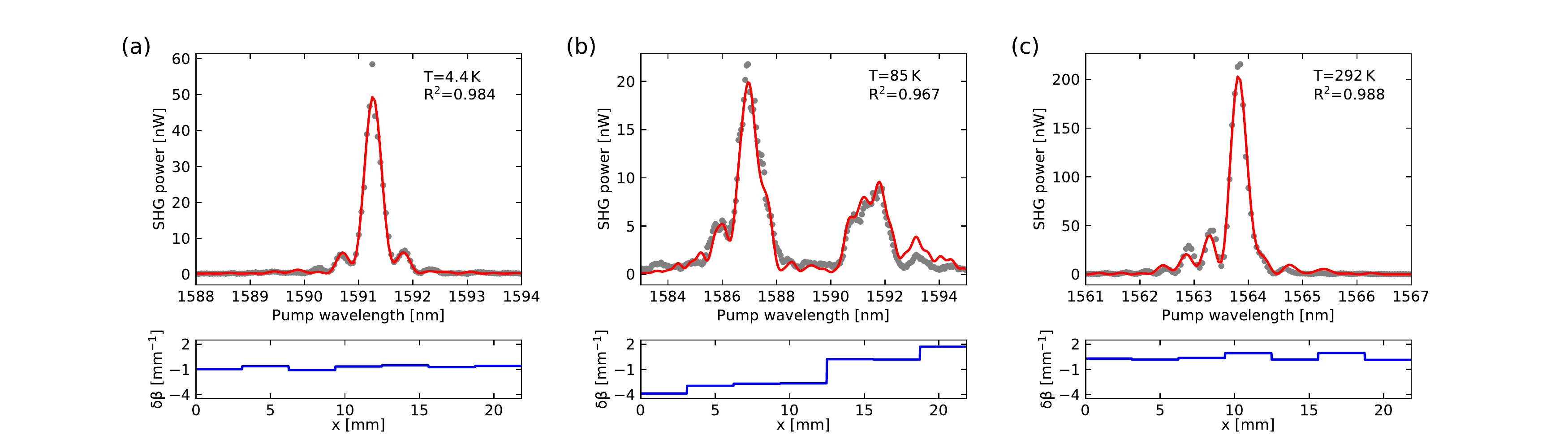}
	\caption{Top: Measured (gray) and fitted (red) SHG spectra at (a) 4.4\,K, (b) 85\,K and (c) 292\,K. Bottom: Corresponding modeled local perturbations $\delta \beta (x)$ in the phase-mismatch arising from the fitted spectra . For details see text.}
	%\caption{(a) Coldest SHG spectrum measured at 4.4\,K with fitted intensity distribution and corresponding modeled perturbation $\delta \beta (x)$ underneath. (b) SHG spectrum with double-peak taken at 85\,K. (c) Room-temperature SHG spectrum measured by the end of the second warm-up.}

	\label{fig:regression_plots}
\end{figure*}
%We observe the phase-matching of the SHG process while 

To measure the effects of temperature change on the SHG, the sample is cooled down to $4.4\,\mathrm{K}$ over $\sim$15 hours and subsequently warmed up over $\sim$20 hours. During temperature cycling, we continuously measure SHG spectra by sweeping the pump wavelength between 1560\,nm and 1595\,nm in steps of 0.1\,nm, over a time period of 260\,s. This corresponds to a maximum temperature variation of 2\,K within one spectrum. Fig.~\ref{fig:experimental_data}(a) shows an example of monitoring the spectra between 4.4~--~292\,K. Two main peaks separated by $\sim$22\,nm arise due to SHG in both the fundamental mode and a higher-order mode. The central wavelength of the SHG peak of both modes shifts towards longer wavelengths with decreasing temperature. 

In Fig.~\ref{fig:experimental_data}(b), the phase-matched wavelength of the fundamental mode is plotted as a function of temperature for four separate measurements, as well as its expected behavior due to the extrapolated Sellmeier equations~\cite{edwards1984,jundt1997}, modified for the waveguide geometry (for details see App.~\ref{chap:refractive_indices}). %As anticipated, the phase-matched wavelength increases towards 0\,K. 
%For all measurement runs, a shift in phase-matched wavelength could be consistently reproduced from 292\,K to 4.4\,K. Across this temperature range, the average measured shift, from $(1562.7\pm1.7)\,\textrm{nm}$ to $(1590.3\pm2.5)\,\textrm{nm}$, is significantly wider than suggested by the extrapolation. 
For all measurement runs, a shift in phase-matched wavelength could be consistently reproduced, from $(1562.7\pm1.7)\,\textrm{nm}$ at 292\,K to $(1590.3\pm2.5)\,\textrm{nm}$ at 4.4\,K. 
The observed shift is significantly larger than suggested by the extrapolation, highlighting the inaccuracy of current models and motivating further investigation in this temperature regime.

%By extrapolating the Sellmeier relations to cryogenic temperatures, we expect the phase-matched wavelength to shift from 1562.9\,nm at 292\,K to 1582.1\,nm at 4.4\,K. 

%We define the SHG conversion efficiency 
%\begin{equation}
%\eta_\mathrm{SHG}=\frac{P_\mathrm{SHG}}{P_\mathrm{pump}^2L^2}\times100\,\%~,
%\end{equation}
%where $P_\mathrm{pump}$ is the pump power coupled into the waveguide and $P_\mathrm{SHG}$ the SHG power generated at the end of the waveguide. 

\section{Transient spectral dynamics}
On top of an underlying reproducible trend in the phase-matched wavelength, we observed significant dynamics in the SHG spectra, occurring on different time scales and at random temperatures, across the four measurements (see Fig.~\ref{fig:experimental_data}). In Fig.~\ref{fig:experimental_data}(a) %(a) shows, as a function of temperature, the SHG spectra we obtained during one warm-up of the sample. 
at least three discontinuities in the phase-matched wavelength can be identified, located around 85\,K, at 180\,K and 249\,K. Between 67\,K and 92\,K, a discontinuity exists over a time scale of 2.5\,h, whereas at 180\,K this takes place over ten minutes, and at 249\,K, this occurs between obtaining two successive spectra, $<$260\,s. Furthermore, experimental data showing even faster discontinuities %at 278\,K 
can be seen in Fig.~\ref{fig:experimental_data}(b), and are discussed in more detail in App.~\ref{chap:supplemental_spectra}.
Since the cooling rate is roughly constant across the relevant temperature range from 50\,K to 250\,K, the different timescales of these phenomena point towards underlying physical processes which modify the phase-matching conditions. Moreover, the spectra shown in Fig.~\ref{fig:experimental_data}(a) might suggest that such perturbations of the phase-matching spectra are mode-dependent, since no discontinuities are observed in the higher-order mode, whilst clearly present in the fundamental mode. However, further experiments demonstrate correlated discontinuities affecting both modes, details of which are provided in App.~\ref{chap:supplemental_spectra}.

Due to the current lack of knowledge of the nonlinear behavior, and relevant material parameters, of lithium niobate under our operating conditions, \textit{a priori} modeling of the measured spectra is at best incredibly challenging.
Therefore, we consider a simplified treatment in which we model a spatially-dependent phase-matching relation along the waveguide. We describe each SHG spectrum by a wavelength-independent perturbation of the phase-mismatch function along the waveguide $\Delta \beta (\lambda_\mathrm{p}) \rightarrow \Delta \beta (\lambda_\mathrm{p}) + \delta \beta (x)$. %This description accounts for perturbations in the refractive indices from all possible origins, without developing a first-principles model of each relevant optical effect, {e.g.} the electro-optic effect, piezoelectric or thermal stress, or a temperature-gradient in the propagation direction. 
The corresponding SHG spectra are calculated by~\cite{helmfrid1993}
\begin{equation}\label{eq:integral}
P_\textrm{SHG}(\lambda_\mathrm{p})\propto \left|\int_0^L e^{-\mathrm{i}\Delta \beta(\lambda_\mathrm{p}) x}\,e^{-\mathrm{i}\int_0^x \delta \beta(x')\mathrm{d}x'}\mathrm{d}x\right|^2~.
\end{equation}
We solve the integral in Eq.~\ref{eq:integral} (using freely available code~\cite{matteos_code}), assuming perturbations $\delta \beta(x)$ consisting of seven sections of equal, fixed width. Seven sections are chosen as a trade-off between computational effort and numerical precision. For each individual SHG spectrum, we combine Monte Carlo simulations with a regression algorithm to determine a perturbation $\delta \beta (x)$ which fits the experimental data, with coefficients of determination $R^2$ ranging from 0.833 to 0.999. A detailed description of the regression method, as well as all determined perturbations from 4.4~--~292\,K, can be found in App.~\ref{chap:regression_app}.

Fig.~\ref{fig:regression_plots} shows three individual spectra (obtained at different temperatures) from Fig.~\ref{fig:experimental_data}, and their corresponding reconstruction. Our simple model matches the measured data remarkably accurately, for ideal spectra such as in Fig.~\ref{fig:regression_plots}(a) at 4.4\,K, as well as the highly distorted profile in Fig.~\ref{fig:regression_plots}(b) at 85\,K and the slightly asymmetric profile at 249\,K in Fig.~\ref{fig:regression_plots}(c). The impressive quality of the fits indicates that the waveguide does indeed undergo localized, temperature-dependent perturbations. 

%During a discontinuity (Fig.~\ref{fig:regression_plots}(b)), such a local perturbation in $\delta\beta(x)$ is significantly larger than the overall variation between 4.4\,K and 292\,K (Figs.~\ref{fig:regression_plots}(a) and (c), respectively). This indicates that the local perturbations can arise, and subsequently stabilize, over timescales dependent on the temperature. 

We attribute the source of these perturbations to the complex interplay of several coupled effects including pyroelectricity, electro-optics and piezo-electric strain \cite{lang1974,jachalke2017} which influence the refractive indices.
The build-up and subsequent relaxation dynamics we observe could be explained by the accumulation and discharge of pyroelectric fields. To support this hypothesis, we consider the magnitudes of these fields which arise during temperature cycling of lithium niobate in vacuum. 
%Without discharging, from 0~~292\,K, we would expect an electric field build-up~\cite{shaldin2008} of the order of $|\Delta E| = \nicefrac{|\Delta P_\mathrm{s}|}{\epsilon_\mathrm{r}\epsilon_0}=45\,\mathrm{kV\,mm}^{-1}$, where $\epsilon_0$ is the electric field constant and $\epsilon_\mathrm{r}$=28 the relative permittivity of lithium niobate along the $c$-axis~\cite{weis1985}. Such fields, larger than the coercive field strength of 21\,kV\,mm$^{-1}$~\cite{gopalan1997}, would likely be sufficient to affect the periodic poling irreversibly.

Without discharging, from 4.4~--~292\,K, we would expect an electric field build-up~\cite{shaldin2008} of the order of $|\Delta E| = \nicefrac{|\Delta P_\mathrm{s}|}{\epsilon_0}=1.3\,\mathrm{MV\,mm}^{-1}$, where $\epsilon_0$ is the electric field constant. Such fields, larger than the coercive field strength of 21\,kV\,mm$^{-1}$~\cite{gopalan1997}, would likely be sufficient to affect the periodic poling irreversibly.

This would result in partial erasing of the poling structure,  corresponding to a permanent increase in the width of the SHG spectrum after temperature cycling, which is not observed. In fact, at both 4.4\,K and 292\,K, we record SHG spectra with the same spectral width of 0.3\,nm~FWHM (see {e.g.} Figs.~\ref{fig:regression_plots}(a) and \ref{fig:regression_plots}(c)). Therefore, we assume some discharge of this field must occur since we are able to reproduce our SHG experiment consistently. 
%This is contradicted by the observation that   This indicates that the dynamics observed during temperature cycling are transient effects%., likely arising from accumulation.
%It is therefore plausible that the pyroelectric fields dissipate via several discharges, occurring at random times. 

Because our sample is in an evacuated cryostat, there exists no efficient pathway for electric charges to dissipate (for example via atmospheric ions). Furthermore, 
%However, 
the long timescales of the dynamics, particularly at cold temperatures, suggest that reduced charge mobility at low temperatures~\cite{akhmadullin1998,dhar2013} plays a role in the discharge dynamics. This hints at discharging pathways through the lithium niobate itself. However, further data on the electrical properties of lithium niobate under these conditions is needed to corroborate this hypothesis.

%The different timescales for the 
%This model is supported by the reduced electric conductivity at colder temperatures, which may explain why the observed dynamics take place on longer timescales when colder.

%Fig.~\ref{fig:regression_plots} shows three spectra measured at different temperatures, with their respective fitted spectra and corresponding perturbations $\delta \beta(x)$. The spectrum in Fig.~\ref{fig:regression_plots}(a) was taken at a temperature of 4.4\,K, with a shape close to an ideal sinc-squared function. Hence the fitted perturbation varies minimally along the waveguide. The largest local perturbations are determined for spectra measured around 85\,K (see Fig.~\ref{fig:regression_plots}(b)). The fitted perturbation varies between $-3.9\,\mathrm{mm}^{-1}$ and 1.7\,$\mathrm{mm}^{-1}$ along the waveguide, equivalent to local perturbations in the poling period from $\delta\Lambda_\mathrm{min}=-23$\,nm to $\delta\Lambda_\mathrm{max}=51$\,nm. This corresponds to a relative variation of the poling period of $(\delta\Lambda_\mathrm{max}-\delta\Lambda_\mathrm{min})/\Lambda=0.8\,\%$. The spectrum plotted in Fig.~\ref{fig:regression_plots}(c) was taken directly after warming up to 292\,K. The asymmetry can be explained by a temperature gradient along the waveguide.

%%%%%%%%%%%%%%%%%
%Across temperature cycles, many SHG spectra deviate significantly from the ideal sinc-squared shape, including the double-peak structure shown in Fig.~\ref{fig:regression_plots}(b), as well as asymmetric SHG spectra as shown in Fig.~\ref{fig:regression_plots}(c).

\section{Conclusion}
In summary, we have demonstrated Type-II SHG down to 4.4\,K in a periodically-poled, titanium in-diffused lithium niobate waveguide. We characterized the shift in the phase-matched wavelength, which is caused by a combination of the temperature-dependent refractive indices, as well as localized changes to the propagation conditions which we attribute to the accumulation and discharge of pyroelectric fields over different timescales. We developed a model which accurately describes SHG spectra deviating from the ideal sinc-squared behavior. Our modeling is a strong indicator that the waveguide properties undergo local perturbations, and provides an indirect way to access dynamics occurring within the waveguide itself.

Our approach not only represents a fundamentally new method to investigate nonlinear optical phenomena, but also kickstarts future nonlinear quantum photonic experiments at low temperature. 
%For some experiments it may be advantageous to mitigate the accumulation of pyroelectric charges. However, 
Below 50\,K, the phase-matched wavelength was consistently reproducible across measurement runs, which indicates that the observed transient dynamics play a minimal role for quasi-phase-matched Type-II SHG at low temperature. Finally, our cryogenic-compatible packaging and characterization allows many more nonlinear optical phenomena to be studied at low temperature, and paves the way for hybrid system integration combining phase-matched processes  with cryogenic quantum technologies.%, such as electro-optic modulation, low-noise frequency conversion and nonclassical state generation.

\section*{Funding Information}
This project is supported by the German Federal Ministry of Education and Research (BMBF) under the
“Quantum Futur” Programme, project number 13N14911, the European Union via the EU quantum flagship project UNIQORN (Grant No. 820474) and the DFG (Deutsche
Forschungsgemeinschaft).

\appendix
\section{Conversion efficiency}\label{sec:conv_eff_app}
\begin{figure}[tbp]
	\centering %<left> <lower> <right> <upper>
		\includegraphics[width=1.0\linewidth,trim={0.8cm 0.3cm 2.4cm 1.6cm},clip]{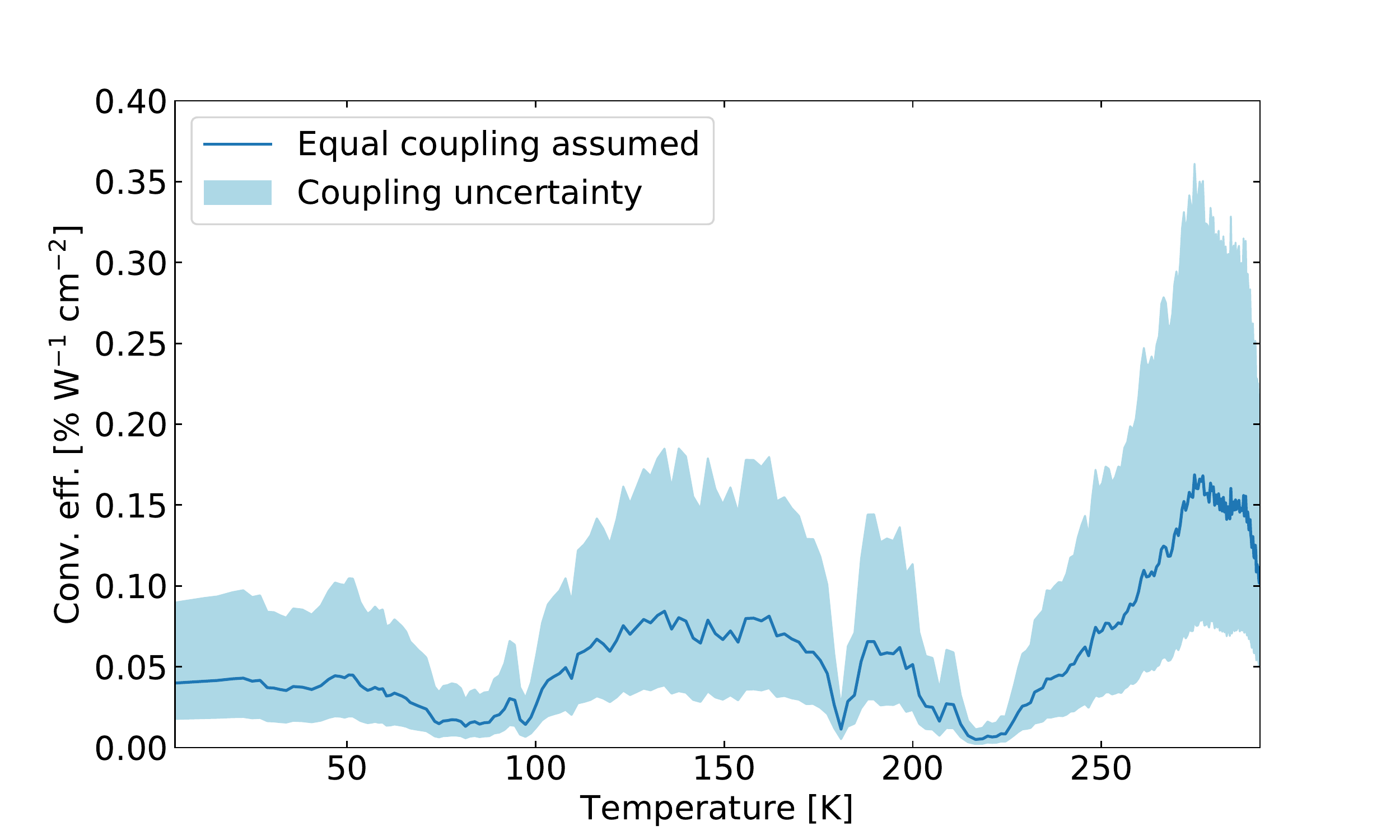}
	\caption{Conversion efficiency as a function of temperature during temperature cycling under the assumption of equal in- and outcoupling (dark blue line), and under the assumptions of perfect incoupling or perfect outcoupling (light blue interval).}
	\label{fig:conversion_efficiency}
\end{figure}

Fig.~\ref{fig:conversion_efficiency} shows the temperature-dependent conversion efficiency for the quasi-phase-matched type II SHG-process which we quantified for the measurement shown in Fig.~2(a) of the main manuscript. We define the conversion efficiency according to
\begin{align}
\eta_\mathrm{SHG}=\frac{P'_\mathrm{SHG}}{{P'_\mathrm{pump}}^2L^2}\times100\,\%,
\end{align}
where $L$=2.5\,cm is the waveguide length and $P'_\mathrm{SHG}$ the amount of SHG power generated with the phase-matched pump wavelength at the end of the waveguide (before outcoupling) . $P'_\mathrm{pump}=\eta_\mathrm{in}P_\mathrm{pump}$ denotes the portion of pump power which is coupled from the input fiber into the waveguide.

Due to thermal contractions while temperature cycling, the incoupling efficiency $\eta_\mathrm{in}$ and outcoupling efficiency $\eta_\mathrm{out}$ vary during the experiment. From our measurements, we know the pump power $P_\mathrm{pump}$ before incoupling and the SHG power after outcoupling $P_\mathrm{SHG}=\eta_\mathrm{out}P'_\mathrm{SHG}$. To determine the quantities $P'_\mathrm{pump}$ and $P'_\mathrm{SHG}$, which are required to calculate the conversion efficiency $\eta_\mathrm{SHG}$, we use the monitored sample transmission
\begin{equation}
T_\mathrm{sample} = \eta_\mathrm{in} \,\eta_\mathrm{out}.
\end{equation}
Taking no Fresnel reflections at the waveguide interconnections into account, we quantify an upper and a lower boundary for the conversion efficiency as a function of temperature (see the light blue area in Fig.~\ref{fig:conversion_efficiency}). For the upper boundary, we assume the best incoupling $\eta_\mathrm{in}=1$ and the worst outcoupling $\eta_\mathrm{out}=T_\mathrm{sample}$, leading to 
\begin{align}
\eta_\mathrm{SHG}^\mathrm{max}=\frac{1}{T}\,\frac{P_\mathrm{SHG}}{P_\mathrm{pump}^2L^2}\times100\,\%~.
\end{align}
Conversely, we assume $\eta_\mathrm{out}=1$ and $\eta_\mathrm{in}=T_\mathrm{sample}$ for the lower boundary of the conversion efficiency, which we quantify by
\begin{align}
\eta_\mathrm{SHG}^\mathrm{min}=\frac{1}{T^2}\,\frac{P_\mathrm{SHG}}{P_\mathrm{pump}^2L^2}\times100\,\%~.
\end{align}
Between these limits, the dark blue line in Fig.~\ref{fig:conversion_efficiency} indicates the conversion efficiency under the assumption of equal coupling, i.e. $\eta_\mathrm{in}=\eta_\mathrm{out}=\sqrt{T_\mathrm{sample}}$,  such that the conversion efficiency is assumed to be
\begin{align}
\overline{\eta}_\mathrm{SHG}=\frac{1}{T^{3/2}}\,\frac{P_\mathrm{SHG}}{P_\mathrm{pump}^2L^2}\times100\,\%~.
\end{align}

\section{Temperature-dependent refractive indices based on modified Sellmeier equations}\label{chap:refractive_indices}
In order to anticipate the temperature-dependence of our SHG process, we calculate temperature- and wavelength-dependent effective refractive indices using a commercial mode solving software based on the finite element method (RSoft FemSIM~\cite{rsoft_femsim}). The effective refractive indices are determined from refractive index cross sections $n_\mathrm{o,e}(y,z,\lambda,T)$, calculated from the bulk refractive indices $n_\mathrm{o,bulk}(\lambda,T)$ and $n_\mathrm{e,bulk}(\lambda,T)$, locally increased by $\delta n_\mathrm{o,e}(\lambda,c)$ due to the indiffusion of titanium:
\begin{align}
n_\mathrm{o}(y,z,\lambda,T)&=n_\mathrm{o,bulk}(\lambda,T)+\delta n_\mathrm{o}(\lambda,c) \nonumber\\
n_\mathrm{e}(y,z,\lambda,T)&=n_\mathrm{e,bulk}(\lambda,T)+\delta n_\mathrm{e}(\lambda,c)
\end{align}
For the bulk refractive indices, we use empirical Sellmeier relations for temperatures above 300\,K, which are extrapolated for our analysis into the regime of cryogenic temperatures. For the ordinary refractive index (for TE polarization), we use ~\cite{edwards1984}:
\begin{align}
n_\mathrm{o,bulk}^2=A_1+\frac{A_2+B_1 H}{\lambda^2 - (A_3+B_2H)^2}+B_3H-A_4\lambda^2
\end{align}
with
\begin{align}
H=(T-24.5)(T+570.5)~,
\end{align}
where $\lambda$ is the wavelength in microns, $T$ the temperature in degrees Celsius and $A_1$,$A_2$,$A_3$,$A_4$,$B_1$,$B_2$ and $B_3$ empirical constants, summarized in Tab.~\ref{tab:sellmeier_constants}.

For the extraordinary refractive index (for TM polarization), we use~\cite{jundt1997}:
\begin{align}
n_\mathrm{e,bulk}^2=a_1+b_1 h + \frac{a_2+b_2 h}{\lambda^2-(a_3+b_3 h)^2}+\frac{a_4+b_4h}{\lambda^2-a_5^2}-a_6\lambda^2
\end{align}
with
\begin{align}
h=(T-24.5)(T+570.82)~,
\end{align}
where $\lambda$ is the wavelength in microns, $T$ the temperature in degrees Celsius and $a_1$,$a_2$,$a_3$,$a_4$,$a_5$,$a_6$,$b_1$,$b_2$,$b_3$ and $b_4$ empirical constants, summarized in Tab.~\ref{tab:sellmeier_constants}.

\begin{table}[t]
  \centering

    \begin{tabular}{rrll}
    \hline
    \multicolumn{2}{l}{Ordinary} & \multicolumn{2}{l}{Extraordinary} \\
    \hline
		%\multicolumn{1}{l}{$H$} & \multicolumn{1}{l}{$(T-24.5)(T+570.5)$} & $h$   & $(T-24.5)(T+570.82)$ \\
    \multicolumn{1}{l}{$A_1$} & \multicolumn{1}{l}{4.9048} & $a_1$ & 5.35583 \\
    \multicolumn{1}{l}{$A_2$} & \multicolumn{1}{l}{0.11775} & $a_2$ & 0.100473 \\
    \multicolumn{1}{l}{$A_3$} & \multicolumn{1}{l}{0.21802} & $a_3$ & 0.20692 \\
    \multicolumn{1}{l}{$A_4$} & \multicolumn{1}{l}{0.027153} & $a_4$ & 100 \\
    \multicolumn{1}{l}{$B_1$} & \multicolumn{1}{l}{$2.2314\times10^{-8}$} & $a_5$ & 11.34927 \\
    \multicolumn{1}{l}{$B_2$} & \multicolumn{1}{l}{$-2.9671\times10^{-8}$} & $a_6$ & $1.5334\times10^{-2}$ \\
    \multicolumn{1}{l}{$B_3$} & \multicolumn{1}{l}{$2.1429\times10^{-8}$} & $b_1$ & $4.629\times10^{-7}$ \\
     &  & $b_2$ & $3.862\times10^{-8}$ \\
          &       & $b_3$ & $-0.89\times10^{-8}$ \\
          &       & $b_4$ & $2.657\times10^{-5}$ \\

    \hline
    \end{tabular}% 
				\caption{Constants in the empiric Sellmeier relations.}

  \label{tab:sellmeier_constants}%
\end{table}%

\begin{figure}[t]
	\centering %l, b, r, t
		\includegraphics[width=0.95\linewidth, trim =8mm 7mm 10mm 8mm, clip]{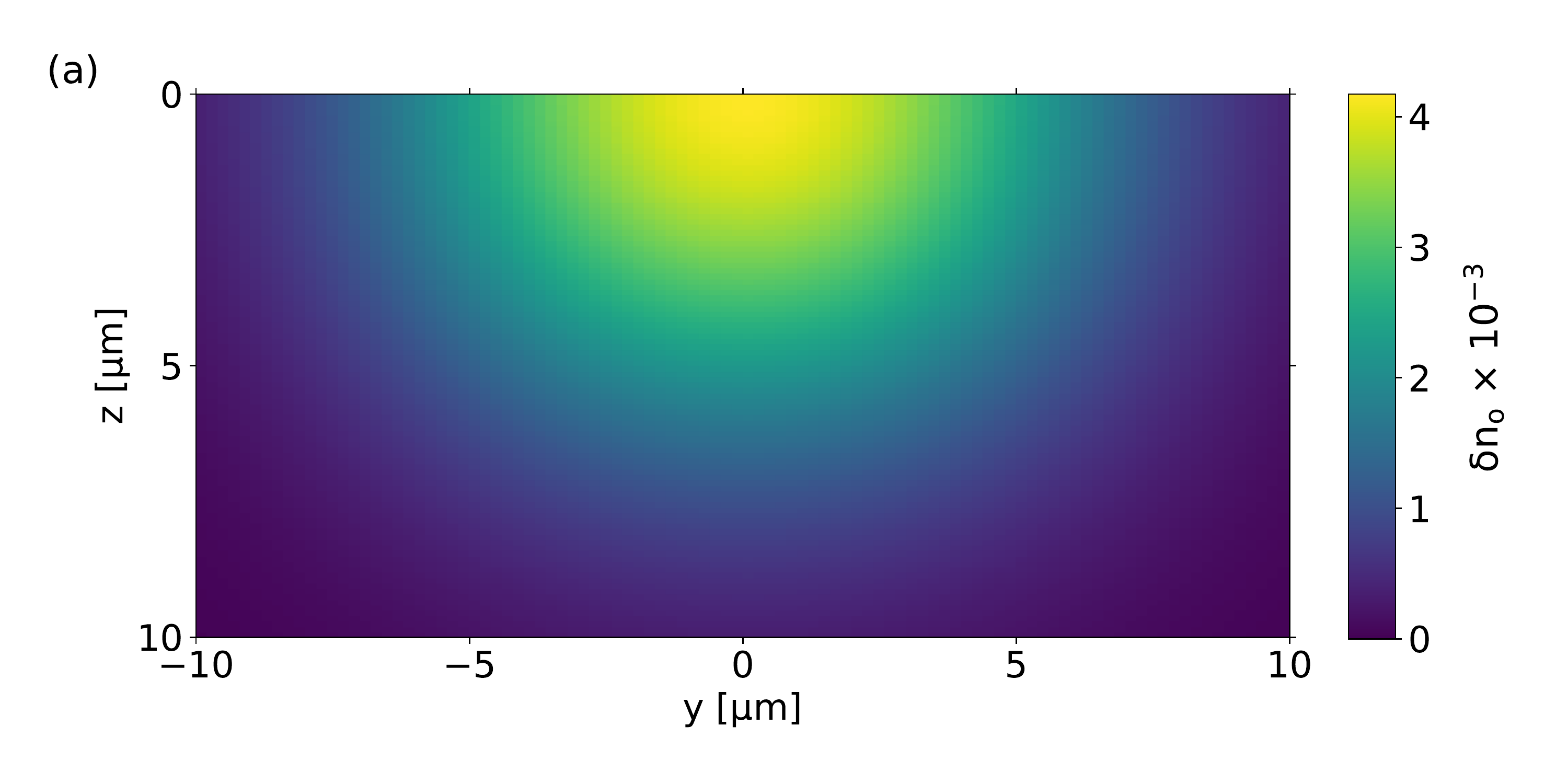}
		\includegraphics[width=0.95\linewidth, trim =8mm 7mm 10mm 8mm, clip]{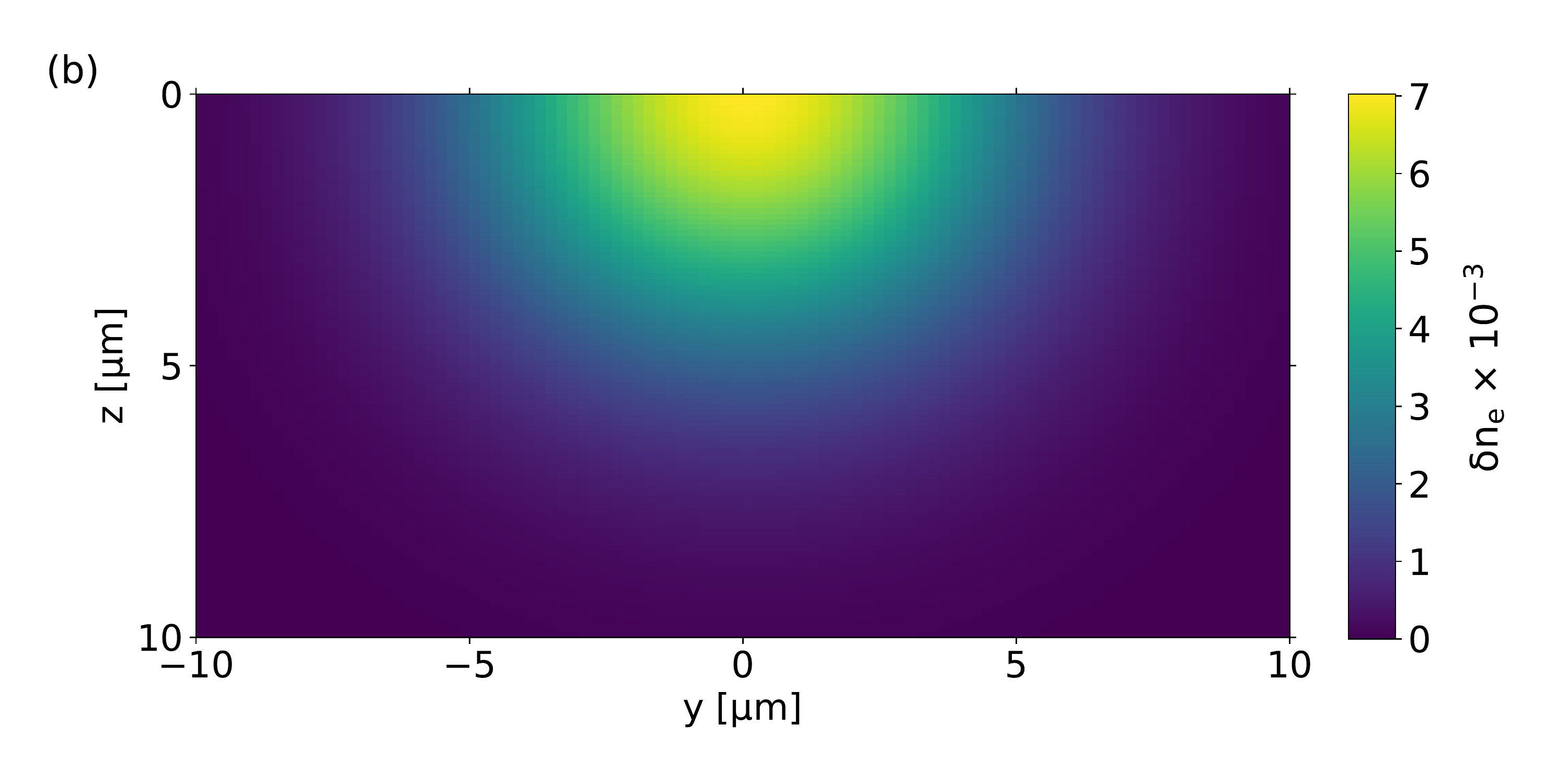}		
	\caption{Modeled increases in the (a) ordinary and (b) extraordinary refractive index of lithium niobate at $\lambda=1.55\,\mathrm{\upmu m}$.}
	\label{fig:index_increase}
\end{figure}

We assume that the spatial refractive index increase due to the indiffusion of titanium is independent of temperature, depending only on wavelength and titanium concentration $c$. We quantify the spatial refractive index increase according to \cite{strake1988} by
\begin{align}
\delta n_\mathrm{o,e}(\lambda,c)=d_\mathrm{o,e}(\lambda) f_\mathrm{o,e}(c)~,
\end{align}
where $d_\mathrm{o,e}(\lambda)$ and $f_\mathrm{o}(c)$ are empirical functions, given by
\begin{align}
d_\mathrm{o}(\lambda)&=\frac{0.67 \lambda^2}{\lambda^2-0.13}~,\\
d_\mathrm{e}(\lambda)&=\frac{0.839 \lambda^2}{\lambda^2-0.0645}~,\\
f_\mathrm{o}(c)&=(Fc)^\gamma~,\\
\textrm{and}\hspace{1cm}f_\mathrm{e}(c)&=E c~, \hspace{2.4cm}%\phantom{aaaaaaaaaaaaaaaa}
\end{align}
with $E=1.2\times10^{-23}\mathrm{cm}^3$, $F=1.3\times10^{-25}\mathrm{cm}^3$ and $\gamma=0.55$.

For the cross-sectional titanium concentration $c(z,y)$, we use the profile for an indiffused titanium strip of thickness $\tau=0.08\,\mathrm{\upmu m}$ and $W=7\,\mathrm{\upmu m}$, specified according to \cite{strake1988} by
\begin{align}
c(z,y)&=c_0 f(u) g(s)
\end{align}
with
\begin{align}
c_0&=\frac{\tau}{a D_y}~,
\end{align}
where $a=1.57\times10^{-23}\,\mathrm{cm}^3$ is as a constant parameter and $D_y$ the diffusion depth perpendicular to the crystal axis. The functions $u$ and $s$ are connected to the crystal coordinates $y$ and $z$ via
\begin{align}
u=\frac{y}{D_y}\hspace{0.8cm}\textrm{and}\hspace{0.8cm}s=\frac{2z}{W}~.
\end{align}
The functions $f(u)$ and $g(s)$ are given by 
\begin{align}
f(u)=\mathrm{exp}(-u^2)
\end{align}
and
\begin{align}
g(s)=\frac{\mathrm{erf}\left[(\nicefrac{W}{2D_z})(1+s)\right]+\mathrm{erf}\left[(\nicefrac{W}{2D_z})(1-s)\right]}{2}
\end{align}
and depend furthermore on the diffusion depth $D_z$ along the crystal axis. The investigated waveguide was fabricated by indiffusing the titanium stripe for 8.5\,h at a temperature of $1060\,^{\circ}\mathrm{C}$, specifying the diffusion depths $D_y=4.0\,\mathrm{\upmu m}$ and $D_z=3.7\,\mathrm{\upmu m}$ determined by our commercial mode solving software \cite{rsoft_femsim,strake1988,korotky1982}. Fig.~\ref{fig:index_increase} shows the resulting modeled cross-sectional refractive index increases at $1.55\,\mathrm{\upmu m}$ wavelength.

\section{Supplemental cryogenic SHG spectra}\label{chap:supplemental_spectra}
\begin{figure}[tbp]
	\centering %l, b, r, t
		\includegraphics[width=\linewidth, trim =1mm 0mm 0mm 3mm, clip]{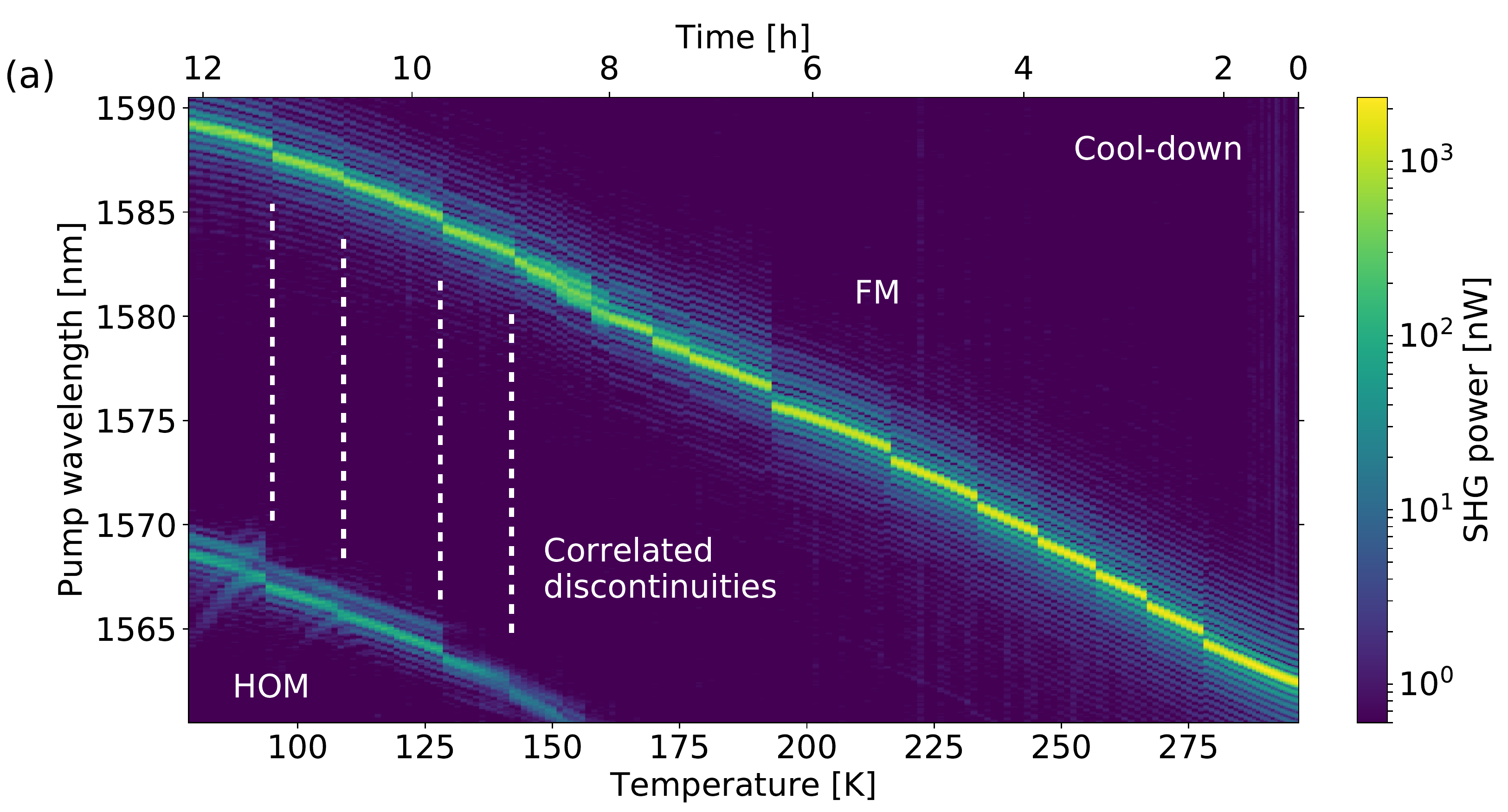}
		\includegraphics[width=\linewidth, trim =1mm 0mm 0mm 3mm, clip]{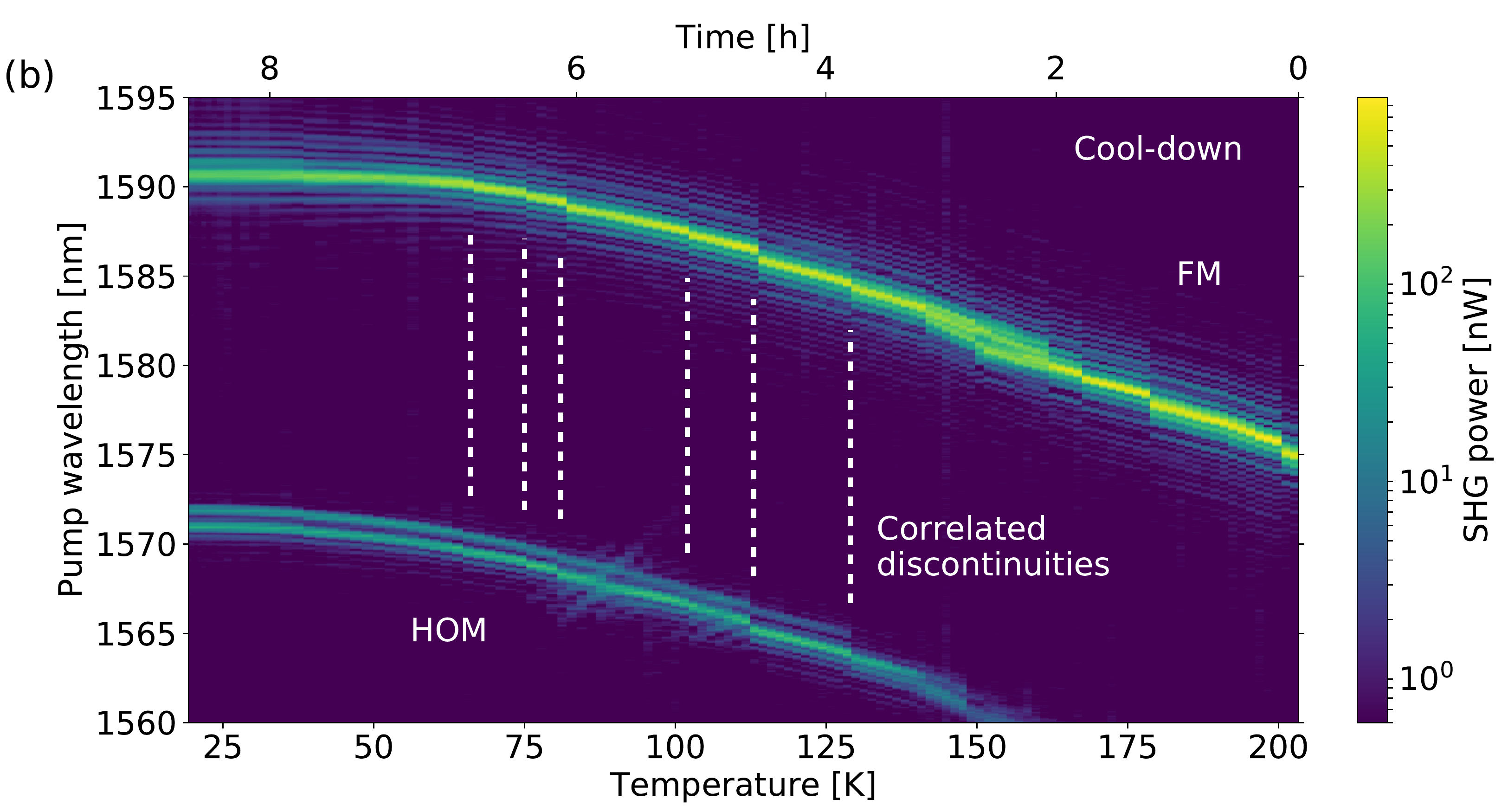}
		\includegraphics[width=\linewidth, trim =1mm 0mm 0mm 3mm, clip]{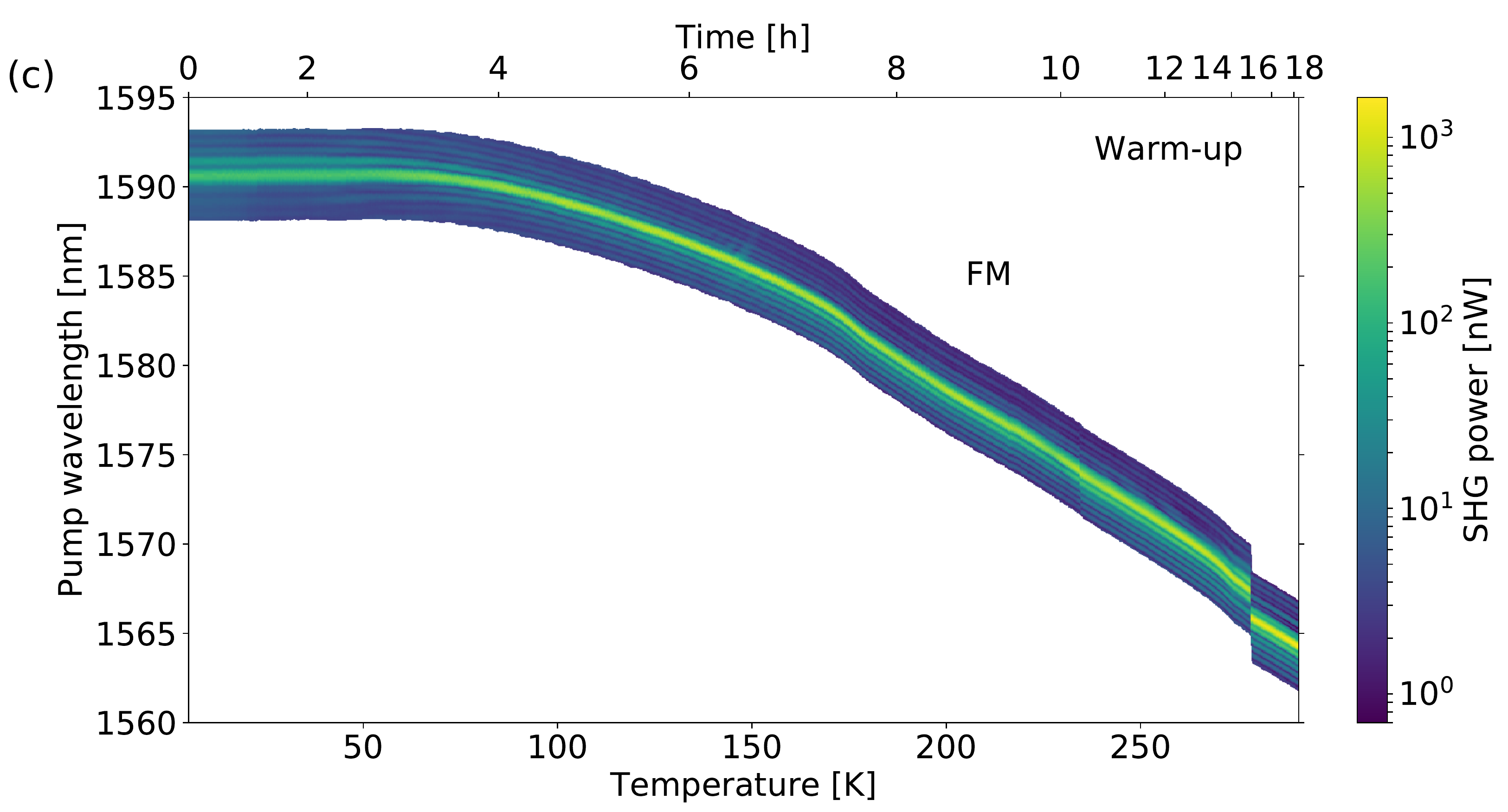}
	\caption{SHG power as a function of pump wavelength and temperature, measured across different temperature and wavelength ranges. The spectra in (a) and (b) show strong correlation of transient dynamics in the fundamental mode (FM) and higher-order mode (HOM). For the measurement in (c), the timing resolution was enhanced by only monitoring SHG in the fundamental mode, resolving a jump in phase-matched wavelength occurring within the sweep of two wavelength increments.}
	\label{fig:experimental_data_sup}
\end{figure}

The monitored SHG spectra summarized in Fig.~2(b) of the main manuscript are shown in Fig.~\ref{fig:experimental_data_sup}. The SHG power was measured as a function of wavelength and temperature across different temperature and wavelength ranges. Fig.~\ref{fig:experimental_data_sup}(a) and Fig.~\ref{fig:experimental_data_sup}(b) show transient dynamics in the SHG spectra, affecting SHG in the fundamental mode simultaneously with SHG in the higher-order mode.

For the data plotted in Fig.~\ref{fig:experimental_data_sup}(c), the pump wavelength was swept over a 5\,nm broad wavelength interval, centered dynamically around the phase-matched wavelength of SHG in the fundamental mode during temperature cycling. This measurement approach allowed us to resolve the discontinuity at 278\,K and see that the SHG spectrum switched within the sweep of two wavelength increments, i.e. within less than 0.7\,s.

\section{Regression of SHG Spectra}\label{chap:regression_app}
We model cryogenic SHG spectra deviating from the ideally expected sinc-squared shape, proceeding in two steps: First we use our experimental data to correct the temperature-dependency of the extrapolation discussed in section~\ref{chap:refractive_indices}. Second, we determine distorted SHG spectra based on perturbations in the phase-mismatch along the waveguide.

\subsection{Temperature-dependent phase-mismatch correction}
The unperturbed phase-mismatch for our Type-II SHG process is given by
\begin{align}
\Delta\beta=2 \pi\left[\frac{\Delta n(\lambda_\mathrm{p},T)}{\lambda_\mathrm{p}}-\frac{1}{\Lambda(T)}\right]~,
\end{align}
depending on the linear combination of (effective) refractive indices
\begin{align}
\Delta n(\lambda_\mathrm{p},T)=2n_\mathrm{TE}(\nicefrac{\lambda_\mathrm{p}}{2},T)-n_\mathrm{TM}(\lambda_\mathrm{p},T)-n_\mathrm{TE}(\lambda_\mathrm{p},T)~.
\end{align}
The linear combination $\Delta n(\lambda_\mathrm{p},T)$ can be approximated based on the extrapolation of the  refractive indices explained in section~\ref{chap:refractive_indices}. However, we observe, that the shift in the phase-matched wavelength is greater than that which is predicted according to the extrapolation. Hence we determine a corrected linear combination $\Delta n^\mathrm{corr}(\lambda_\mathrm{p},T)$ by adding a corrective function $a(T)$ to the extrapolated linear combination $\Delta n^\mathrm{extrp}$:
\begin{align}
\Delta n^\mathrm{corr}(\lambda_\mathrm{p},T)=\Delta n^\mathrm{extrp}+a(T)
\end{align}
At every temperature $T$, we determine $a(T)$ from our measured cryogenic SHG spectra, based on the average phase-matched pump wavelength $\overline{\lambda}_\mathrm{p}$ at which the SHG signal is maximal. At the phase-matched wavelength, $\Delta \beta=0$, and we can calculate $a(T)$ from
\begin{align}
\frac{\Delta n^\mathrm{extrp}(\overline{\lambda}_\mathrm{p},T)+a(T)}{\overline{\lambda}_\mathrm{p}}-\frac{1}{\Lambda (T)}=0~.
\end{align}
From this, we use the phase-mismatch $\Delta\beta^\mathrm{corr}$, calculated from the corrected linear combination $\Delta n^\mathrm{corr}(\lambda_\mathrm{p},T)$:
\begin{align}
\Delta\beta^\mathrm{corr}=2 \pi\left[\frac{\Delta n^\mathrm{corr}(\lambda_\mathrm{p},T)}{\lambda_\mathrm{p}}-\frac{1}{\Lambda(T)}\right]
\end{align}

\subsection{Locally perturbed phase-matching}
To describe SHG spectra deviating from the ideal sinc-squared shape, we introduce local perturbations in the (corrected) phase-mismatch along the waveguide

\begin{equation}
\Delta \beta^\mathrm{corr} (\lambda_\mathrm{p}) \rightarrow \Delta \beta ^\mathrm{corr}(\lambda_\mathrm{p}) + \delta \beta (x).
\end{equation}
From such perturbations, the shape of the SHG spectrum can be theoretically calculated by \cite{helmfrid1993}
\begin{equation}\label{eq:integral_sup}
P_\textrm{theo}(\lambda_\mathrm{p})\propto \left|\int_0^L e^{-\mathrm{i}\Delta \beta^\mathrm{corr}(\lambda_\mathrm{p}) x}\,e^{-\mathrm{i}\int_0^x \delta \beta(x')\mathrm{d}x'}\mathrm{d}x\right|^2~.
\end{equation}
To compare theoretical SHG spectra with experimental data, we normalize for our analysis theoretically modeled and experimentally measured power distributions according to
\begin{align}
p_\textrm{theo}(\lambda_\mathrm{p})=\frac{P_\textrm{theo}(\lambda_\mathrm{p})}{\int P_\textrm{theo}(\lambda_\mathrm{p}) \mathrm{d}\lambda_\mathrm{p}}
\end{align}
and
\begin{align}
p_\textrm{exp}(\lambda_\mathrm{p})=\frac{P_\textrm{exp}(\lambda_\mathrm{p})}{\int P_\textrm{exp}(\lambda_\mathrm{p}) \mathrm{d}\lambda_\mathrm{p}}~.
\end{align}

With a regression approach, we determine perturbations in the phase-mismatch $\delta \beta (x)$ resulting in modeled SHG spectra $p_\textrm{theo}(\lambda_\mathrm{p})$ which fit the experimental SHG spectra $p_\textrm{exp}(\lambda_\mathrm{p})$. Since we consider SHG in the waveguide ground mode, the side peak due to SHG in the higher mode is cut off the experimental spectra.

For all SHG spectra $p_\textrm{exp}(\lambda_\mathrm{p})$ measured at temperatures $T$, we determined a perturbation $\delta \beta (x)$ generating the respective intensity distribution. For this, we approximated the perturbations in a general manner by a step function consisting of seven equally broad plateaus of heights $c_i$, while choosing seven steps as a trade-off between fitting quality and computational effort:
\begin{align}
\delta \beta_\mathrm{step} (x,c_1,\cdots,c_7)=c_i \hspace{5mm}\mathrm{for}&\hspace{5mm} \frac{i-1}{7}\,L \leq y < \frac{i}{7}\,L, \nonumber \\
\mathrm{where}&\hspace{5mm}i=1,\cdots,7
\end{align}

For each measured SHG spectrum, we fit an intensity distribution $p_\textrm{model} (\lambda_\mathrm{p},L,c_1,\cdots,c_7)$ based on a seven-step perturbation $\delta \beta_\mathrm{step} (x,c_1,\cdots,c_7)$. We solve the integral in Eq.~\ref{eq:integral_sup} using Simpson's method implemented in the ``PyNumericalPhasematchingv1.0b'' package \cite{matteos_code}. The sample length was fitted once at $T$=4.4\,K and kept constant during the further analysis since perturbations along the waveguide have no influence on this parameter. The determined length is $L$=21.85\,mm, which can be interpreted as an effective length being shorter than the actual sample length of 25\,mm due to waveguide imperfections.

\begin{figure}[tbp]
	\centering %l, b, r, t
		\includegraphics[width=\linewidth, trim =8mm 7mm 25mm 15mm, clip]{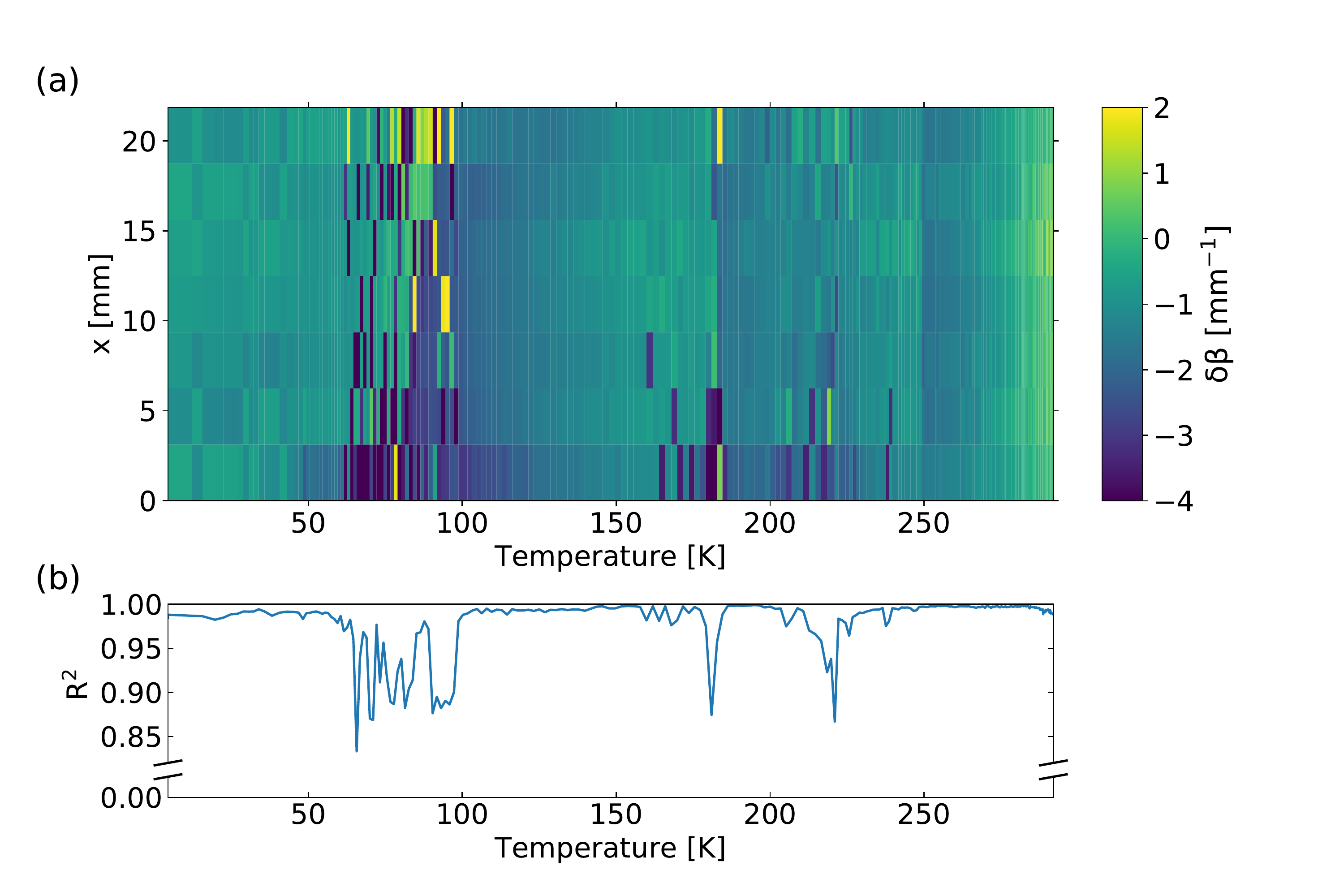}
	\caption{Fitted perturbations $\delta n$ along the waveguide in the $x$-direction as a function of temperature $T$. The coefficient of determination $R^2$ indicates that the modeled SHG spectra agree very well with the measured SHG spectra exhibiting a single SHG peak. Spectra with double phase-matching could be modeled by refractive index perturbations as well.}
	\label{fig:regression_map}
\end{figure}

For each spectrum we performed a two-step regression procedure with respect to the parameters $\{c_i\}$. In the first step, $10^3$~spectra based on random regression parameters are simulated. In a second step, the best random guess $\{c_i\}_\mathrm{init}$ leading to the least residual sum of squares (RSS) is used as initial point for a function regression with the Levenberg-Marquardt algorithm~\cite{scipy2020}. In Fig.~\ref{fig:regression_map}(a), the determined perturbations $\delta n(x)$ along the waveguide are plotted as a function of temperature $T$. The coefficient of determination $R^2$, plotted as a function of temperature in Fig.~\ref{fig:regression_map}(b), varies between 0.833 and 0.999.

\bibliography{sample}

\end{document}